\documentclass{aastex63}
\usepackage{graphicx,natbib}
\usepackage[percent]{overpic}
\usepackage{graphicx,subfigure,amsmath}
\usepackage{hyperref}
\usepackage{lineno}
\newcommand\gsim{\,\lower3pt\hbox{$\sim$}\llap{\raise2pt\hbox{$>$}}\,}
\newcommand\lsim{\,\lower3pt\hbox{$\sim$}\llap{\raise2pt\hbox{$<$}}\,}

\shorttitle{Simulation of 13-Dec-2006 CME}
\shortauthors{Fan}

\begin{document}

\title{An improved MHD simulation of the 2006 December 13 coronal mass ejection of active region NOAA 10930}

\correspondingauthor{Yuhong Fan}
\email{yfan@ucar.edu}

\author{Yuhong Fan}
\affil{High Altitude Observatory, National Center for Atmospheric Research, \\
3080 Center Green Drive, Boulder, CO 80301, USA}

\begin{abstract}
We present a magnetohydrodynamic (MHD) simulation of the coronal mass ejection (CME) on 13 December 2006 in the emerging $\delta$-sunspot active region 10930, improving upon a previous simulation by Fan (2016) as follows. (1) Incorporate an ambient solar wind instead of using a static potential magnetic field extrapolation as the initial state. (2) In addition to imposing the emergence of a twisted flux rope, also impose at the lower boundary a random electric field that represents the effect of turbulent convection, which drives field-line braiding and produces resistive and viscous heating in the corona. With the inclusion of this heating, which depends on the magnetic field topology, we are able to model the synthetic soft X-ray images that would be observed by the X-Ray Telescope (XRT) of the Hinode satellite, produced by the simulated coronal magnetic field. We find that the simulated pre-eruption magnetic field with the build up of a twisted magnetic flux rope, produces synthetic soft X-ray emission that shows qualitatively similar morphology as that observed by the Hinode/XRT for both the ambient coronal loops of the active region and the central inverse-S shaped ``sigmoid” that sharpens just before the onset of the eruption. The synthetic post-flare loop brightening also shows similar morphology as that seen in the Hinode/XRT image during the impulsive phase of the eruption. It is found that the kinematics of the erupting flux rope is significantly affected by the open magnetic fields and fast solar wind streams adjacent to the active region.

\end{abstract}


\section{Introduction}
Significant progress has been made in theory and numerical simulations to
understand the basic magnetic configurations and mechanisms for the
onset of major solar eruptions such as eruptive flares and coronal
mass ejections \citep[e.g. reviews by][]{Forbes:etal:2006,
Chen:2011, Green:etal:2018}.
Recently, MHD simulations that use observed data to construct
the initial state and the boundary driving conditions are being developed
and are playing an increasingly important role in studying the realistic
complex magnetic field evolution in observed solar eruptive events
\citep[e.g.][]{Jiang:etal:2013, Kliem:etal:2013, Inoue:etal:2014,
Amari:etal:2014, Fan:2016, Hayashi:etal:2018, Toeroek:etal:2018,
Guo:etal:2019, Liu:etal:2022}.

\citet{Fan:2011, Fan:2016} and \citet{Amari:etal:2014} have carried out
MHD simulations to model the magnetic field evolution of the X3.4 flare
and the associated halo CME that occurred on December 13, 2006 in
active region (AR) 10930. The photospheric magnetic field evolution of
AR 10930 was characterized by an emerging positive
polarity spot against the southern edge of a dominant pre-existing negative
spot. The emerging positive spot showed substantial counter-clockwise
rotation (by $540 ^{\circ}$) and eastward motion as it
grew \citep{Min:Chae:2009}, suggesting the emergence of a twisted magnetic
flux rope.  Motivated by this observation, \citet{Fan:2016} modeled the
pre-eruption magnetic field by imposing at the lower boundary the
(partial) emergence of an east-west oriented magnetic torus into an
initial potential coronal magnetic field extrapolated from a smoothed
magnetogram from the Solar and Heliospheric Observatory (SOHO)/Michelson
Doppler Imager (MDI) about 6 hours before the flare.
An east-west oriented magnetic flux rope with more than 1 wind of
total twist is built up quasi-statically in the corona and subsequently
erupts to produce a CME in the MHD simulation.  The simulated magnetic
field evolution is able to qualitatively explain several observed
features of the event seen in the images by the X-Ray Telescope (XRT)
on board the Hinode satellite, including the
formation of an X-ray sigmoid brightening just before the onset of the
eruption and the morphology of the post-flare loop brightening during
the impulsive phase of the flare.
\citet{Amari:etal:2014} modeled the pre-eruption magnetic field evolution
by constructing a sequence of force-free field solutions extrapolated from
 the observed photosphere vector magnetograms from the Solar Optical
Telescope (SOT) of Hinode. They also found the
formation of a twisted magnetic flux rope with more than 1 wind of twist
in the pre-eruption magnetic field just before the onset of the observed
major eruption.  They further carried out an MHD simulation of the
subsequent dynamic evolution using the force-free field solution as
the initial state, driven with the appropriated lower boundary
conditions, and found the loss of equilibrium of the flux rope
that produced a CME.

For all the above models, the comparisons with observations have
been limited to identifying the morphology of selected field lines with
certain features in the observables such as the soft X-ray emission observed
by the Hinode/XRT. Also for the above simulations, the initial ambient corona
is modeled with a static, potential magnetic field configuration, without
considering the effect of the solar wind outflows.
In this work, we improve upon the simulation of \cite{Fan:2016} by incorporating
a more explicit treatment of the background coronal heating that represents the
effect of turbulent convection and varies self-consistently based on the
formation of the strong current layers in the 3D coronal magnetic field.
With this heating, we are able to model the synthetic soft X-ray emission as
would be observed by the Hinode/XRT produced by the simulated coronal magnetic
field of the active region.  This allows a more direct comparison with the
observation.  With this background coronal heating, we also initialize a
partially open coronal magnetic field with a background solar wind as the initial
state for our simulation of the event.
In section \ref{sec:model} we describe the MHD model and the setup of the
improved simulation. The result of the simulation is shown in section
\ref{sec:results} and a summary and discussion is given in section \ref{sec:summary}

\section{Model Description}
\label{sec:model}
In this simulation, we use the ``Magnetic Flux Eruption'' (MFE) code to solve
the set of semi-relativistic MHD equations as described in \citet{Fan:2017}.
Compared to the previous simulation of \citet{Fan:2016}, a more sophisticated treatment of
the thermodynamics is used where the internal energy equation explicitly takes into
account the non-adiabatic effects in the corona, including the field-aligned thermal
conduction, optically thin radiative cooling, and coronal heating.
The readers are referred to \citet{Fan:2017} for the MHD equations being solved and a
description of the numerical methods. Here we only describe the modifications
made in the current numerical model relative to \citet{Fan:2017}.
For the heating term in the internal energy equation
\citep[term H in eq. (5) of][]{Fan:2017},
we no longer include the empirical coronal heating
\citep[eq. (14) in][]{Fan:2017},
but only include the heating resulting from the dissipation of the
kinetic and magnetic energies due to the numerical diffusion effects
\citep[see the description in the
first paragraph on p.3 of][]{Fan:2017}. 
As described later in this section, for this simulation,
we impose at the lower boundary a random
electric field representing the effect of turbulent convection that
drives field line braiding, and the resultant (numerical) resistive
and viscous dissipation of the magnetic and kinetic energies provides
the necessary heating of the corona.
The heating varies spatially and temporally self-consistently with
the locations of the strong current layers that form in the 3D
magnetic field.

\begin{figure}[htb!]
\centering
\includegraphics[width=0.8\textwidth]{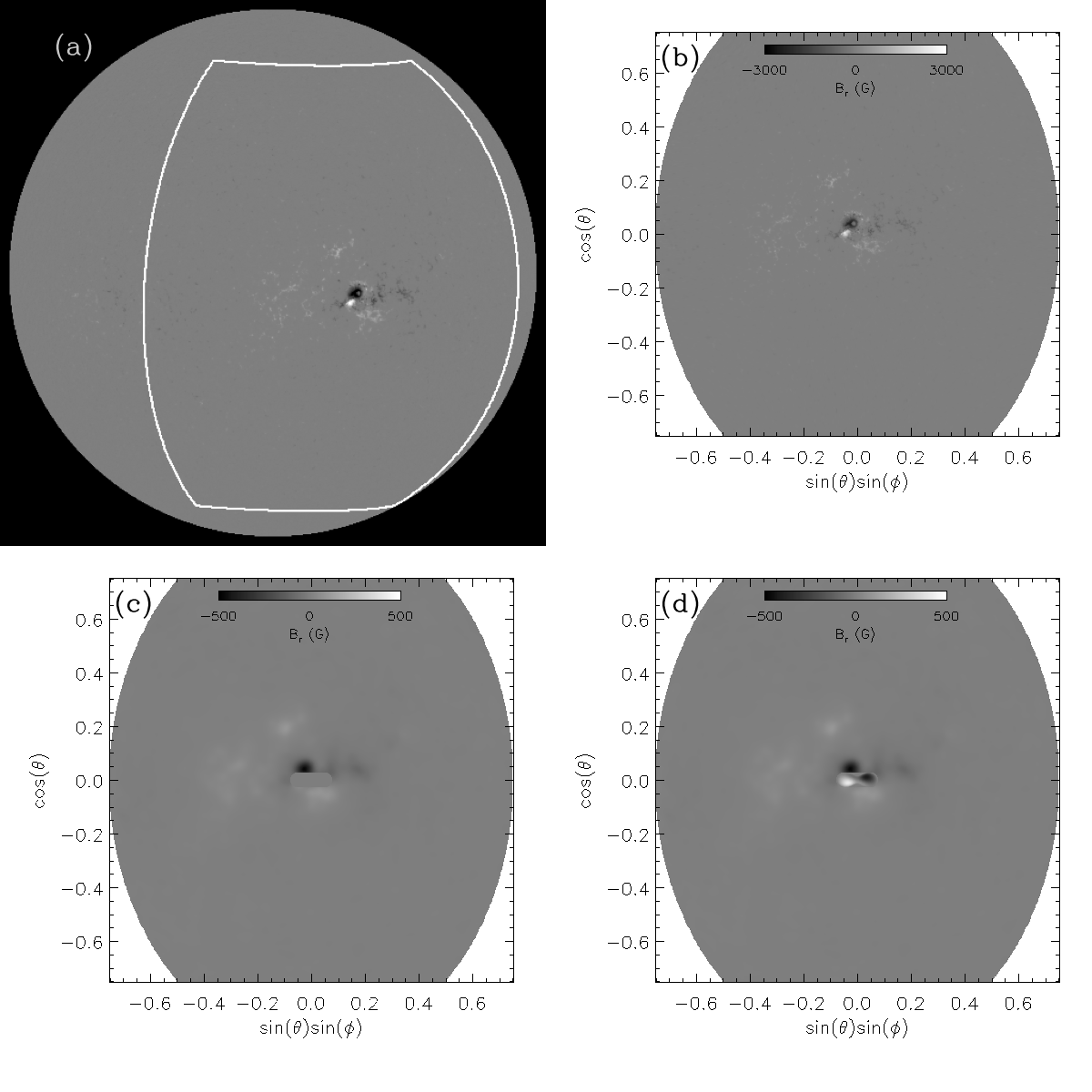}
\caption{(a) SOHO MDI full-disk magnetogram at 20:51:01 UT on 2006
December 12. The white box encloses the area to be used as the lower boundary
of the simulation domain. (b) Radial magnetic field $B_r$ in the region
enclosed by the white box in (a) as viewed straight on at the center of the
emerging active region. (c) $B_r$ of the extrapolated potential field
(extrapolated from the $B_r$ in (b))
in the horizontal slice at 14 Mm height and with the field in a central
region zeroed out, where the emergence of a twisted magnetic torus is to be
imposed. This $B_r$ is the initial lower boundary normal magnetic field for
the simulation. (d) $B_r$ on the lower boundary at the time of the impulsive
phase of the eruption in the simulation.}
\label{fig:init}
\end{figure}

\begin{figure}
\centering
\includegraphics[width=0.5\textwidth]{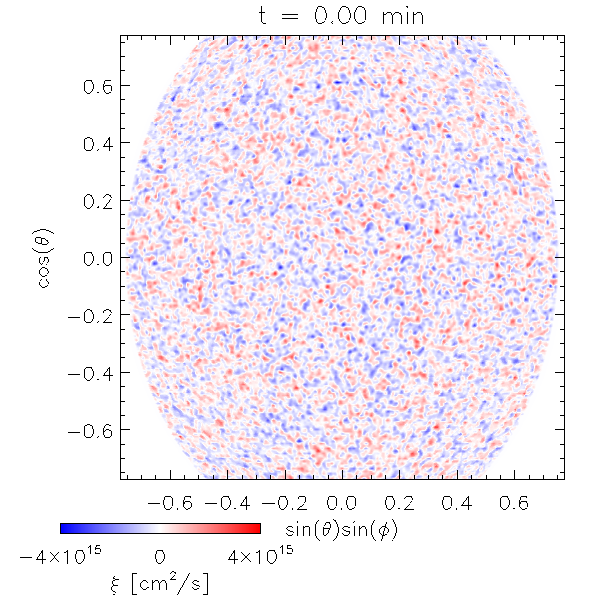}
\caption{A snapshot of the $\xi$ field used in specifying
$\bf{E}_{h}^{\rm{random}}$ in equation (\ref{eq:xi}).
An animated version of the figure is available, which shows the
evolution of the random cellular pattern of the $\xi$ field for about 5 hours.}
\label{fig:xi}
\end{figure}

We set up the initial state with a background solar wind as follows.
The white boxed region on the MDI full-disk magnetogram
at 20:51:01 UT on December 12 (Figure \ref{fig:init}(a)) is extracted
as the lower boundary of the spherical wedge simulation domain, for which the
the emerging active region is centered at $\theta=0$ and $\phi=0$ in the
simulation spherical coordinates as shown in Figure \ref{fig:init}(b).
The radial extent of the simulation domain is from $1 R_{\odot}$
to $10 R_{\odot}$, where $R_{\odot}$ is the solar radius. The
latitudinal and longitudinal widths of the simulation domain
are respectively $117.2^{\circ}$ and $98.3^{\circ}$, centered on the
active region.  The spherical wedge simulation domain is resolved
with a grid of $832(r) \times 480(\theta) \times 504(\phi)$ that is
stretched in all three directions such that the highest grid resolution
($1.03$ Mm) is near the lower boundary and centered on the active region.
A potential field extrapolation from the photosphere normal magnetic field
at the domain's lower boundary shown in Figure \ref{fig:init}(b) is carried out.
Then the radial magnetic field in the horizontal plane at 14 Mm height of
the extrapolated potential field, with a central area corresponding to the
region of the observed flux emergence zeroed out, is used as the initial lower
boundary normal magnetic field (Figure \ref{fig:init}(c)) for the simulation.
It is considered as the pre-emergence normal magnetic field at the base
of the corona. We then initialize the domain with a potential field extrapolated
from this lower boundary normal magnetic field (shown in Figure \ref{fig:init}(c))
together with a hydrostatic atmosphere, and numerically evolve the coronal
domain driven at the lower boundary with the following random, horizontal
electric field $\bf{E}_{h}^{\rm{random}}$:
\begin{equation}
c {\bf{E}}_{h}^{\rm{random}} = - \nabla_{h} ( \xi B_r ) .
\label{eq:xi}
\end{equation}
In the above $c$ denotes the speed of light, the subscript $h$ denotes the
horizontal directions perpendicular
to the $r$ (radial) direction, $B_r$ is the normal magnetic field in the lower
boundary and $\xi$ is a time dependent field that is made up of a superposition
of 62402 randomly placed cells of opposite sign values. Each of the cells is
spatially a 2D Gaussian profile with a size scale of 5.22 Mm and a central
amplitude of $8.14 \times 10^{14} \rm{cm^2/s}$, and temporally
a sinusoidal function with a period and life time of 11.89 min.
A snapshot of the $\xi$ field is illustrated in Figure \ref{fig:xi} with a movie
showing the evolution of the time-dependent $\xi$ field in the online
version of the paper.
$\bf{E}_{h}^{\rm{random}}$ effectively drives random rotations
of the footpoints of the $B_r$ flux concentrations, with positive (negative)
$\xi$ corresponding to clockwise (counter-clockwise) rotation.
Note, since $\bf{E}_{h}^{\rm{random}}$ is given as the gradient of a potential
(eq. [\ref{eq:xi}]), it does not produce any change of $B_r$ on the surface
and therefore preserves the (observed) normal magnetic flux distribution.
Here with the use of the staggered grid and the constrained transport
scheme \citep[e.g.][]{Stone:Norman:1992:b}, the zero-circulation nature of
$\bf{E}_{h}^{\rm{random}}$ and hence the zero change of $B_r$ is enforced
numerically to machine precision at the lower boundary surface.
$\bf{E}_{h}^{\rm{random}}$ represents the effect of turbulent convection that
drives field line braiding and produces (numerical) resistive and viscous heating
in the corona.
Many past numerical studies have used similar ways of driving coronal heating by
imposing random motions representative of the turbulent convection
at the photospheric lower boundary of the MHD simulations
\citep[e.g.][]{Gudiksen:etal:2005, Warnecke:Peter:2019}.

We note that the $\nabla_{h} ( \xi B_r )$ used for the random electric
field in equation (\ref{eq:xi}) is of the same form as the
$\nabla_s ( \zeta B_n )$ used for the lower boundary electric field in the
STITCH (STatistical InjecTion of Condensed Helicity) subgrid model
described in \citet[][see eq. (1) in that paper]{Dahlin:etal:2021},
with $\nabla_{h}$ corresponding to $\nabla_{s}$, $\xi$ to $\zeta$,
and $B_r$ to $B_n$.  The difference is that a large-scale, like-signed $\zeta$
field is used in the STITCH electric field. In contrast, the random electric
field here uses a collection of small-scale cells of random signs for the
$\xi$ field.
Thus here the random electric field represents the driving effect of a
small-scale turbulent convection with no net helicity transported into the corona.
On the other hand the STITCH electric field represents a large-scale statistical
mean effect of a helical convection that builds up net helicity in the corona
\citep{Mackay:etal:2014, Dahlin:etal:2021}.

The treatment of the thermodynamic conditions at the lower boundary
is the same as that described in \citep[][see eqs. (17)(18) and the
associated text in that paper]{Fan:2017}.
The boundary conditions for the side and the top
boundaries are also the same as \citet{Fan:2017}.

\begin{figure}[htb!]
\centering
\includegraphics[width=0.7\textwidth]{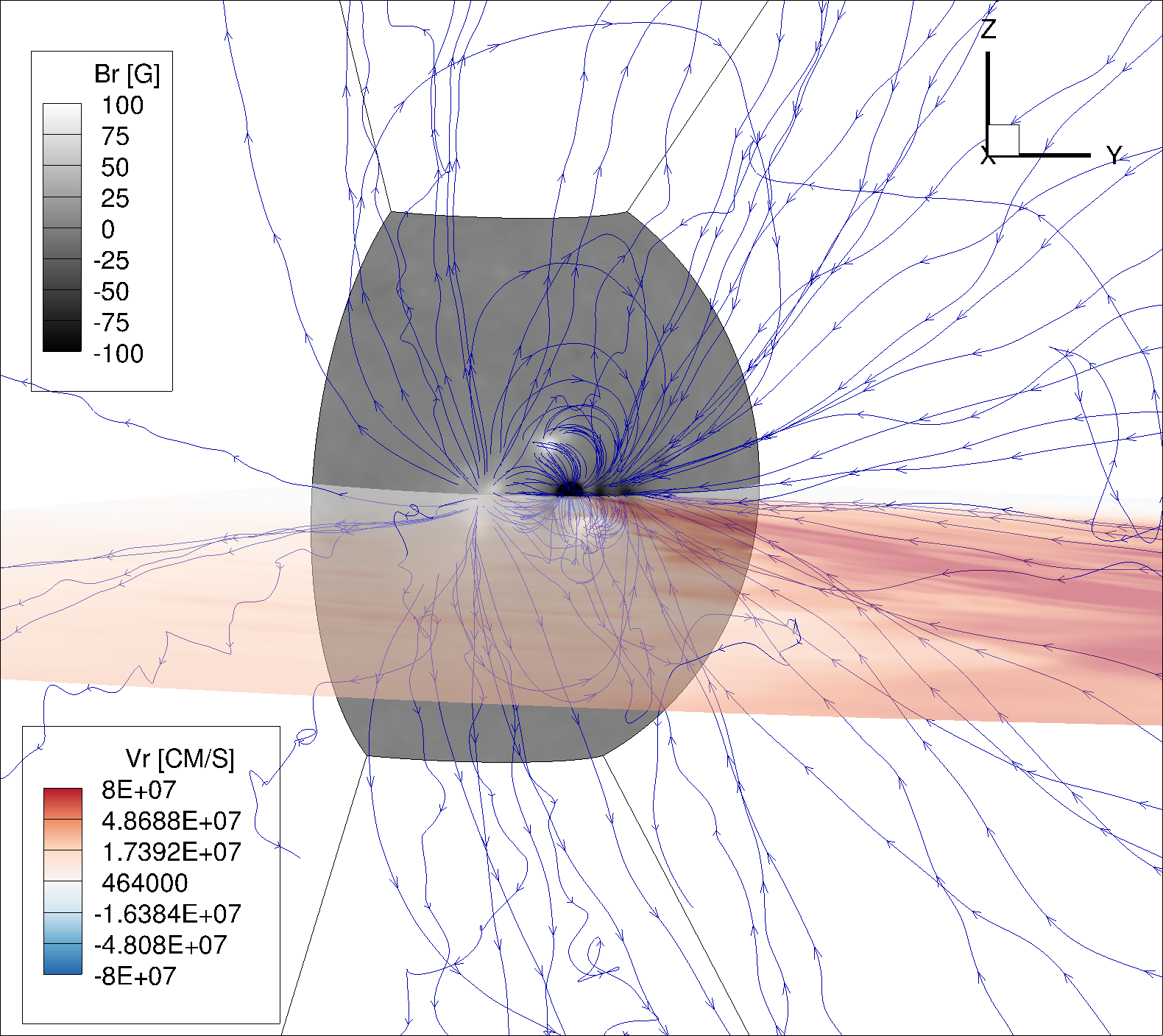}
\caption{The quasi-steady initial state containing a partially open coronal
magnetic field with a background solar wind. Blue lines are magnetic field
lines with arrows marking the magnetic field directions.
A $r$-$\phi$ slice at the latitude of the dominant pre-existing
negative sunspot of the active region is shown with color
indicating the radial velocity. The gray-scale colored bottom boundary surface
shows the normal magnetic field $B_r$.
The open magnetic field lines show large wiggles and kinks, which are due to the
the large amplitude Alfv{\'e}n waves driven by the random electric field at the
lower boundary.}
\label{fig:windinit}
\end{figure}

We numerically evolve the corona domain driven with the above
random electric field at the lower boundary until it reaches a quasi-steady
partially open coronal magnetic field with a background solar wind as shown
in Figure \ref{fig:windinit}.
This is then used as the initial state for the
subsequent simulation of the build up and eruption of the CME magnetic
field driven by flux emergence at the lower boundary.
For the initial state, we see low-latitude open coronal-hole magnetic fields
adjacent to the closed fields of the active region on both its east and
west sides.
There are strong solar wind outflows emanating
from the dominant pre-existing negative sunspot and the adjacent region to
the west of the spot,
as can be seen from the (red) radial velocity $v_r$ in the $r$-$\phi$
slice extending from the active region.
The open magnetic field lines show large wiggles and kinks, which correspond
to the large amplitude Alfv{\'e}n waves driven by the random electric field
at the lower boundary.

\begin{figure}[htb!]
\centering
\includegraphics[width=0.7\textwidth]{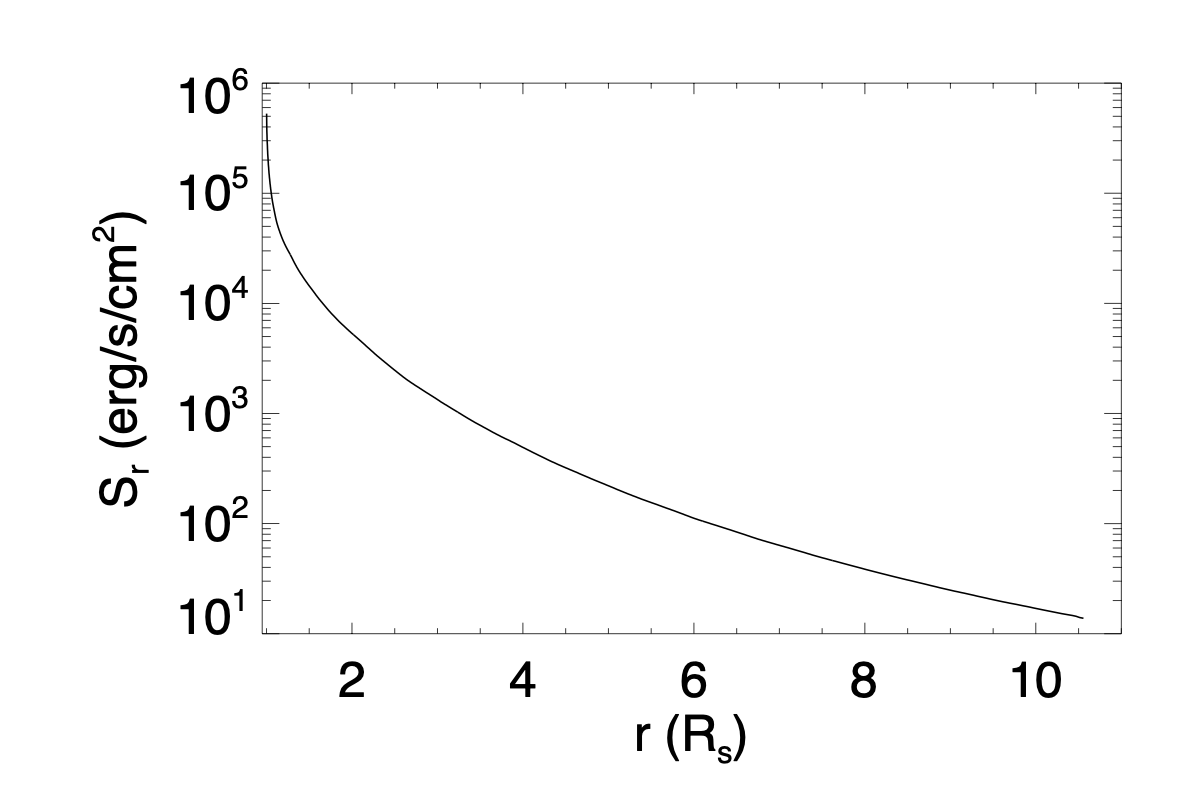}
\caption{Radial profile of the time and horizontally averaged Poynting flux
density (see text for the expression of $S_r$) for the quasi-steady initial state.}
\label{fig:windinit_poyntflux}
\end{figure}

\begin{figure}[htb!]
\centering
\includegraphics[width=0.5\textwidth]{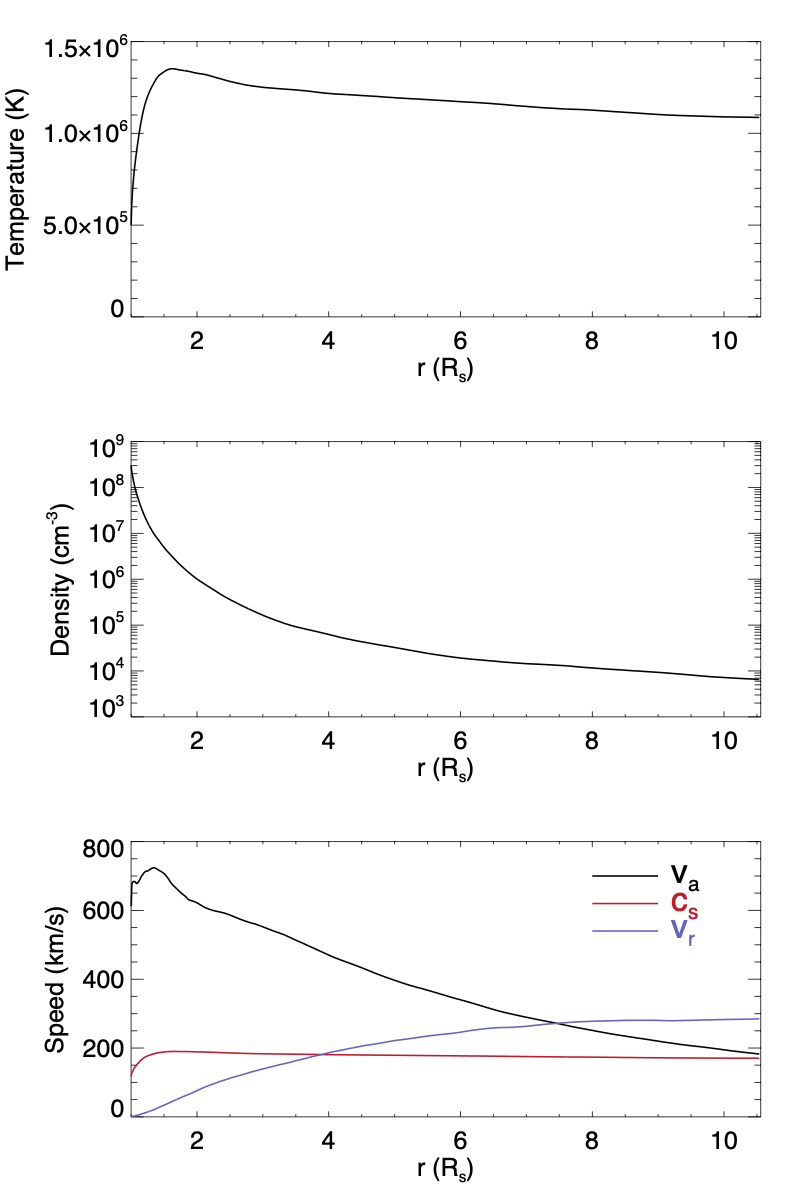}
\caption{Radial profiles of the horizontally averaged temperature (top),
density (middle), and Alfv{\'e}n speed, sound speed, and radial
flow speed (bottom) of the quasi-steady initial state.}
\label{fig:1dprofs}
\end{figure}

The random electric field imposed at the lower boundary drives a Poynting
flux through the coronal domain,
whose time and horizontally averaged radial profile of the flux density
$S_r = \left < (c/4\pi) ({\bf E} \times {\bf B})_r \right >$ is
shown in Figure \ref{fig:windinit_poyntflux}, where $<>$ represents averaging
over spherical shells and over time (for about 10 hours in the quasi-steady
state).
The divergence of the Poynting flux provides the energy input for heating the
corona and accelerating the solar wind \citep[e.g.][]{Rempel:2017}.
The averaged upward Poynting flux density at the lower boundary
(base of the corona) is found to be $5.3 \times 10^5 {\rm erg/s/cm^2}$,
which is consistent with the necessary energy flux to sustain the heating of
the quite-sun corona \citep[e.g.][]{Withbroe:1988,Rempel:2017}.
The r.m.s. velocity near the lower boundary is found to be about 13 km/s, compared to
the mean sound speed of 117 km/s and the mean Alfv{\'e}n speed of 615 km/s.
Thus the velocities near the lower boundary resulting from the random electric field
are very sub-sonic and sub-alfv{\'e}nic.
Figure \ref{fig:1dprofs} shows the radial profiles of the horizontally
averaged temperature, density, Alfv{\'e}n speed, sound speed, and radial
flow speed of the quasi-steady initial state.
It can be seen that reasonable mean temperature, density profiles
of the corona are maintained.  
The (horizontally averaged) mean Alfv{\'e}n speed significantly
exceeds the mean sound
speed for most of the height range of the domain. The mean
radial outflow speed exceeds the mean sound speed at about
4 solar radii and exceeds the mean Alfv{\'e}n speed at about
7 solar radii.

With the above numerically evolved quasi-steady state as the
initial state, we then impose at the lower boundary, in
addition to the above random electric field, the emergence
of an east-west oriented twisted magnetic torus in the zeroed-out
region with the electric field ${\bf E}_{h}^{\rm{emg}}$:
\begin{equation}
c {\bf E}_{h}^{\rm{emg}} = - ( {\bf v}_{\rm rise} \times {\bf B}_{\rm tube})_h ,
\end{equation}
that corresponds to the rise of a magnetic torus ${\bf B}_{\rm tube}$
at a constant velocity ${\bf v}_{\rm rise}$ through the lower
boundary, as was done in \citet{Fan:2016}.
The expressions for ${\bf B}_{\rm tube}$ and ${\bf v}_{\rm rise}$
are given in \citet[][see eqs. (4)-(7) and the associated text in that
paper]{Fan:2016} with the following changed values of the parameters for
the torus: the minor radius of the torus is $a=0.028 R_{\odot}$ and the major
radius is $R'=0.0504 R_{\odot}$, the rate of field-line twist about
the curved axis of the torus is $q/a= 0.102 \, {\rm rad/Mm}$, the field
strength at the curved axis of the torus is $B_t a /R' = 361 {\rm G}$,
the center of the torus is initially located at
($ r_c = 0.922 R_{\odot}, \theta_c = 90^{\circ}, \phi_c = 0^{\circ}$)
in the simulation spherical coordinate system and is assumed to be rising at
a constant velocity ${\bf v}_{\rm rise} = v_{\rm rise} {\hat {\bf r}}_c$
with $v_{\rm rise} = 3.9 \rm{km/s}$.
In this simulation, the pre-emergence lower boundary (base of the corona)
magnetic field (see Figure \ref{fig:init}(c)) is obtained from the extrapolated
potential field at 14 Mm height, and has a significantly stronger
peak strength (peaks at about $500$ G) than that used for the
lower boundary (peaks at about $200$ G) in the simulation of
\citet{Fan:2016}, which is obtained by heavily smoothing the photosphere
magnetogram. Because of this, in this simulation we also use a significantly
stronger, more compact magnetic torus for the imposed flux emergence, as given
by the above changed parameters, compared to that used in the \citet{Fan:2016}.
At the end of the imposed flux emergence in the simulation,
the total emerged (unsigned)
flux is approximately equal to the total flux originally in the zeroed out
area in Figure \ref{fig:init}(c).
In summary, in this simulation, we impose at
the lower boundary ${\bf E}_{h}^{\rm{emg}} + {\bf{E}}_{h}^{\rm{random}}$,
where ${\bf E}_{h}^{\rm{emg}}$ drives the emergence of a twisted magnetic
torus to build up the twisted pre-eruption coronal magnetic field, while
the random electric field ${\bf{E}}_{h}^{\rm{random}}$
continues to drive the background corona heating.

\section{Simulation Results}
\label{sec:results}

\begin{figure}[htb!]
\centering
\includegraphics[width=0.9\textwidth]{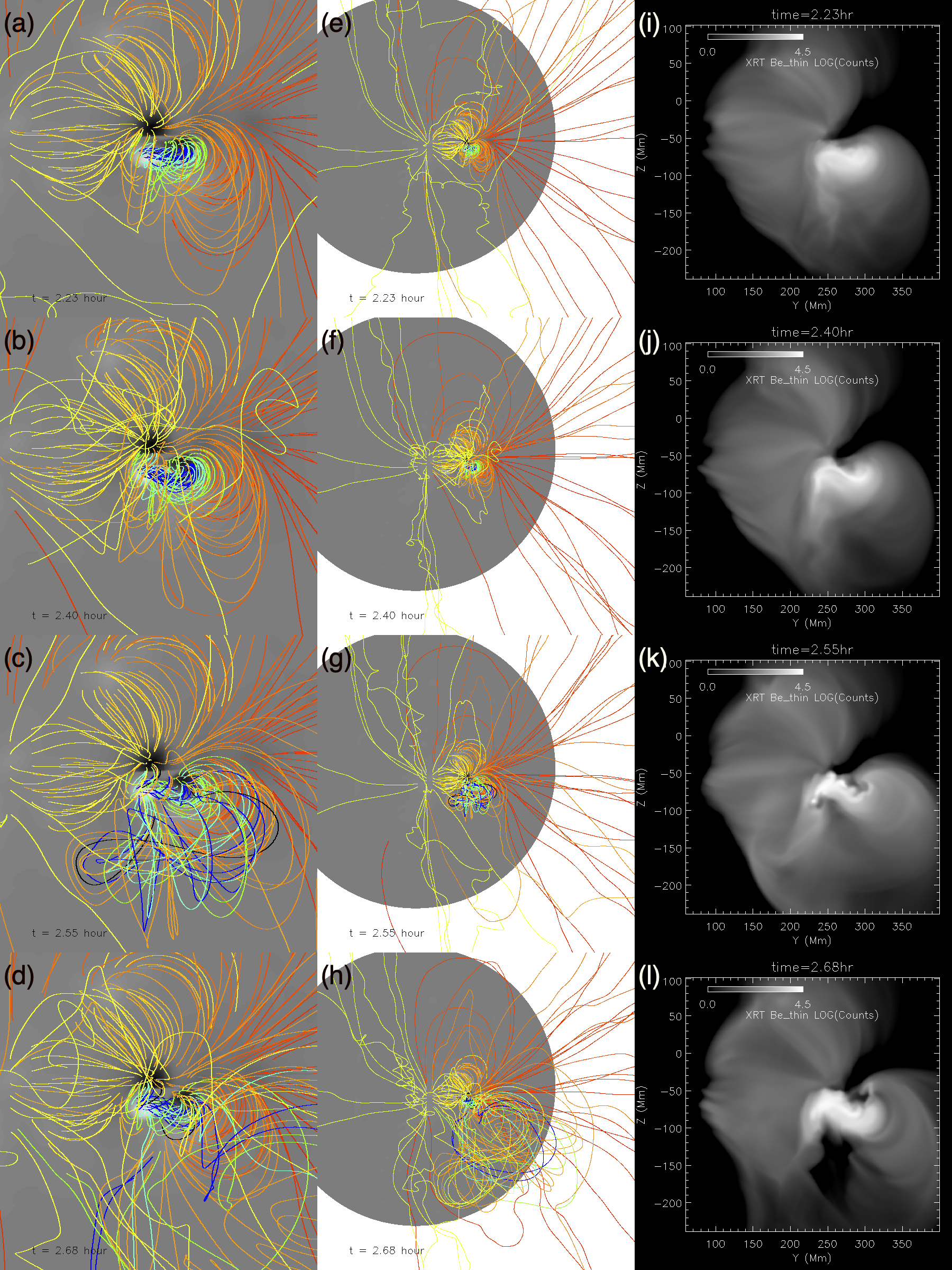}
\caption{Snapshots showing the evolution of the 3D magnetic field lines
(left column: a zoomed-in view, middle column: a far view) and the synthetic
Hinode/XRT images (right column). An animated version of this figure is
available, which shows the entire $3.96$ hours of the simulated evolution from
the beginning of the driving flux emergence through the eruption.
In the animated synthetic X-ray images, it is easier to identify an outgoing
band of diffuse brightening that emanates from the sigmoid at about $t= 2.45$
hour and moves southward until leaving the field of view at about $t=2.58$ hour,
which corresponds to the hot core of the erupting flux rope as discussed
in the text.}
\label{fig:evol_fdl_xrt}
\end{figure}

Figure \ref{fig:evol_fdl_xrt} shows a sequence of snapshots of the evolution
of the 3D magnetic field lines (left column a zoomed-in view, middle column
a far view) and the synthetic Hinode/XRT images (right column) produced by
the simulation.
The synthetic XRT images are computed by integrating along the line of sight
(LOS) as viewed from earth, through the simulation domain:
\begin{equation}
I_{\rm XRT} = \int n_e^2 (l) f_{\rm Be\_thin} (T(l)) \, dl,
\label{eq:xrtsynth}
\end{equation}
where $l$ denotes the length along the LOS through the simulation
domain, $I_{\rm XRT}$ denotes the modeled intensity in XRT ${\rm Be\_thin}$ channel
(shown in the right column images in Figure \ref{fig:evol_fdl_xrt} in LOG scale), 
$n_e$ is the electron number density, and $f_{\rm Be\_thin} (T(l))$ is the
instrument temperature response function for the ${\rm Be\_thin}$ channel obtained
by calling {\tt make\_xrt\_wave\_resp} and
{\tt make\_xrt\_temp\_resp} functions in the IDL SolarSoft XRT package.

\begin{figure}[htb!]
\centering
\includegraphics[width=0.55\textwidth]{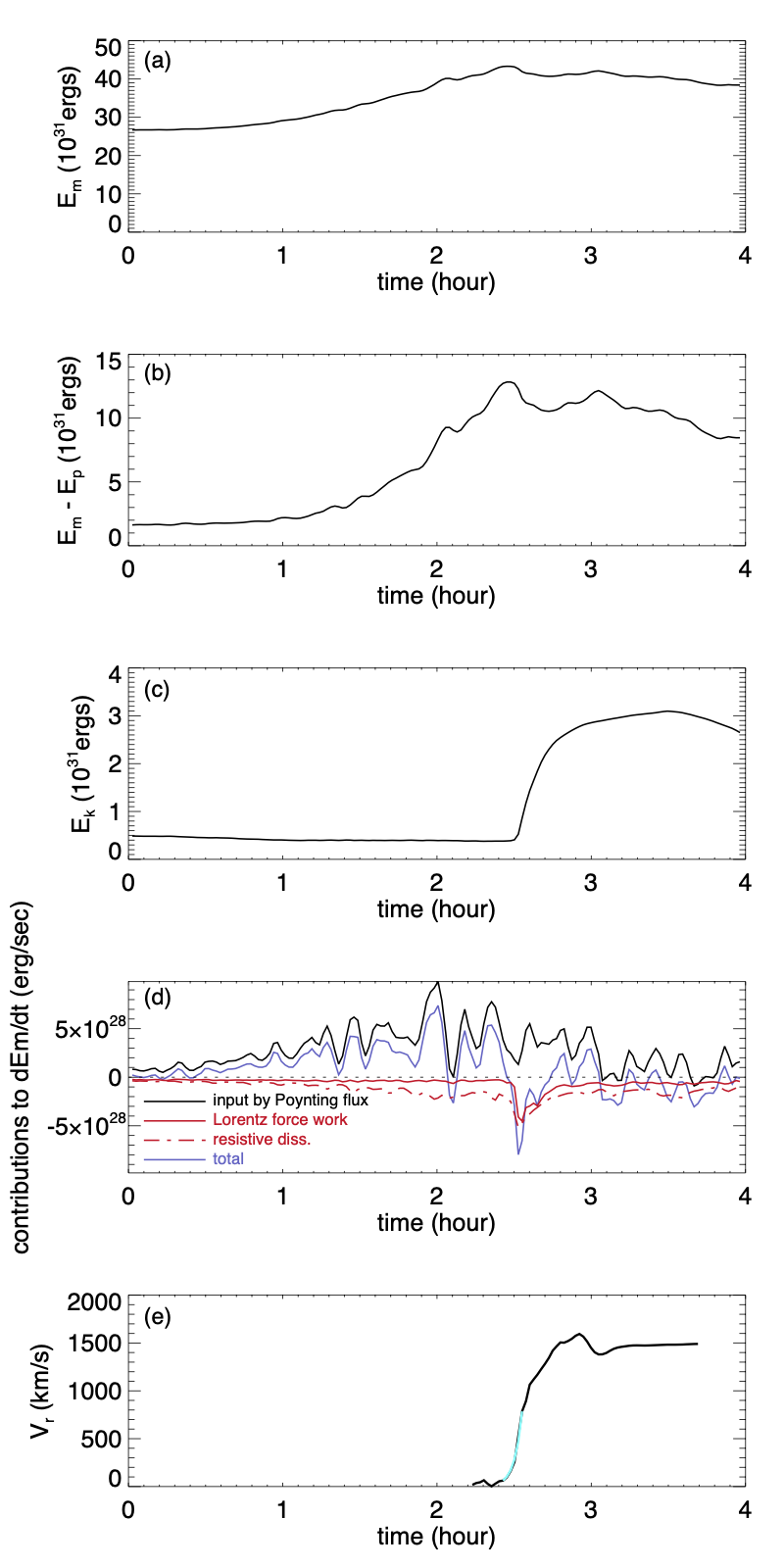}
\caption{(a) shows the evolution of the total magnetic energy $E_m$; (b)
shows the evolution of the free magnetic energy $E_m - E_p$, where $E_p$
is the corresponding potential field energy; (c) shows the evolution of
the total kinetic energy $E_k$; (d) shows the contributions to the rate of change
of the total magnetic energy ($dE_m/dt$) due to: the input of magnetic
energy by the Poynting flux integrated over the boundaries (black curve), the
release of magnetic energy resulting from the Lorentz force work (solid red
curve), the release of magnetic energy by the numerical resistive
dissipation (red dashed-dotted curve), and the sum of the above source
and sinks (blue curve); (e) shows the evolution of the radial velocity
of a Lagrangian tracer point placed in the core field region of the
emerged flux rope just before the onset of the eruption (see the green star
point in Figure \ref{fig:alpha_bpdecay}) and tracked in the velocity field. The cyan curve
is a fitted function of an exponential growth of the radial velocity:
$v_r = v_0 \exp((t-t_0)/{\tau})$ over the period from time $t_0 = 2.43$ hour to
time $t_1 = 2.55$ hour, with $\tau = 176$ sec and $v_0 = 62$ km/s.}
\label{fig:emep_ek_demdt_vr_evol}
\end{figure}

Due to the imposed flux emergence by ${\bf E}_{h}^{\rm{emg}}$ at the lower
boundary, a twisted magnetic flux rope is built up quasi-statically in
the corona over the time period from $t=0$ to about $t=2.4 \, {\rm hr}$,
and an inverse-S shaped sigmoid brightening develops in the synthetic XRT
images (see the two top rows in Figure \ref{fig:evol_fdl_xrt} and the associated
movie).
During this time period the total magnetic energy ($E_m$) and the free magnetic
energy ($E_{\rm m} - E_{\rm p}$, where $E_{\rm p}$ is the corresponding
potential field magnetic energy) show an overall increase as can be seen in
Figures \ref{fig:emep_ek_demdt_vr_evol}(a)(b) (for $t=0$ to roughly
$t=2.4 \, {\rm hr}$),
while the emerging flux rope remains confined by the closed field of the
active region with the kinetic energy maintaining roughly steady at the level of
the ambient solar wind (Figure \ref{fig:emep_ek_demdt_vr_evol}(c)).
In Figure \ref{fig:emep_ek_demdt_vr_evol}(d), we show the various
contributions to the rate of change
of the total magnetic energy ($dE_m/dt$) due to: the input of magnetic
energy by the Poynting flux integrated over the boundaries (black curve), the
release of magnetic energy resulting from the Lorentz force work (solid red
curve), the release of magnetic energy by the numerical resistive
dissipation (red dashed-dotted curve), and the sum of the above source
and sinks (blue curve).
During the quasi-static phase (from $t=0$ to about $t=2.4 \, {\rm hr}$), we
see that the energy input by the Poynting flux integrated over the boundaries
is for most of the time greater than the sum of the total magnetic energy
release in the domain by the Lorentz force work and the numerical resistive
dissipation, and thus the magnetic energy builds up over this period.
The magnetic energy release due to the Lorentz force work is generally
smaller than that due to the numerical resistive dissipation in this
simulation. We see that the numerical resistive dissipation increases
gradually during the quasi-static phase due to the flux emergence.
On the other hand the magnetic energy release due to the Lorentz
force work remains relatively small because the emerged flux rope remains
confined in a quasi-static state.

At about $t=2.5 \, {\rm hr}$ there is a sharp increase of the magnetic
energy release by the Lorentz force work (red solid curve in
Figure \ref{fig:emep_ek_demdt_vr_evol}(d)), which coincides with a
rapid acceleration of the flux rope
(see Figure \ref{fig:emep_ek_demdt_vr_evol}(e))
and a rapid increase of the total kinetic energy
(Figure \ref{fig:emep_ek_demdt_vr_evol}(c)), indicating the loss of
equilibrium and eruption of the flux rope.
Both the Lorentz force work and the numerical resistive dissipation show
a sharp increase (red curves in Figure \ref{fig:emep_ek_demdt_vr_evol}(d)),
resulting in a relatively sharp decrease in the
magnetic energy (Figures \ref{fig:emep_ek_demdt_vr_evol}(a)(b),
at about $t=2.5 \, {\rm hr}$).
After this sudden release of the magnetic energy, the magnetic energy
gradually turns to increasing with time again
(Figures \ref{fig:emep_ek_demdt_vr_evol}(a)(b)) due to the continued
magnetic flux emergence imposed at the lower boundary, until about
$t=3 \, {\rm hr}$ when we stop the imposed flux emergence.
The snapshots in Figures \ref{fig:evol_fdl_xrt}(c)(g)
(at $t=2.55 \, {\rm hr}$) and
Figures \ref{fig:evol_fdl_xrt}(d)(h) (at $t=2.68 \, {\rm hr}$)
show the morphology of some of the erupting twisted field lines.
By tracking a Lagrangian tracer placed in the core field region of the
emerged flux rope just before the onset of the eruption, we find that its
rise speed (Figure \ref{fig:emep_ek_demdt_vr_evol}(e)) undergoes an
exponential growth starting from about $t= 2.43 \, {\rm hr}$ to about
$t=2.55 \, {\rm hr}$ (see the cyan curve),
and it eventually accelerate to a final steady speed of about
as $1500$ km/s for this tracer point in the erupting flux rope. Thus the
time for the onset of eruption of the flux rope is about $t=2.43$ hour,
corresponding to the start of the exponential growth of the rise speed.
It reaches a peak acceleration of about
$3 \, {\rm km/s^2}$ at $t=2.53 \, {\rm hr}$. This peak
accleration is in the high end range of the observed peak accelerations of
CMEs \citep[e.g.][]{Bein:etal:2011}. The high field strength
($\sim 400$ G) of the emerged flux rope at the base of the corona in
this simulation certainly contributes to the fast acceleration of the
eruption \citep[e.g.][]{Green:etal:2018}.

\subsection{The pre-eruption magnetic field and the initiation of eruption}
\begin{figure}[htb!]
\centering
\includegraphics[width=0.8\textwidth]{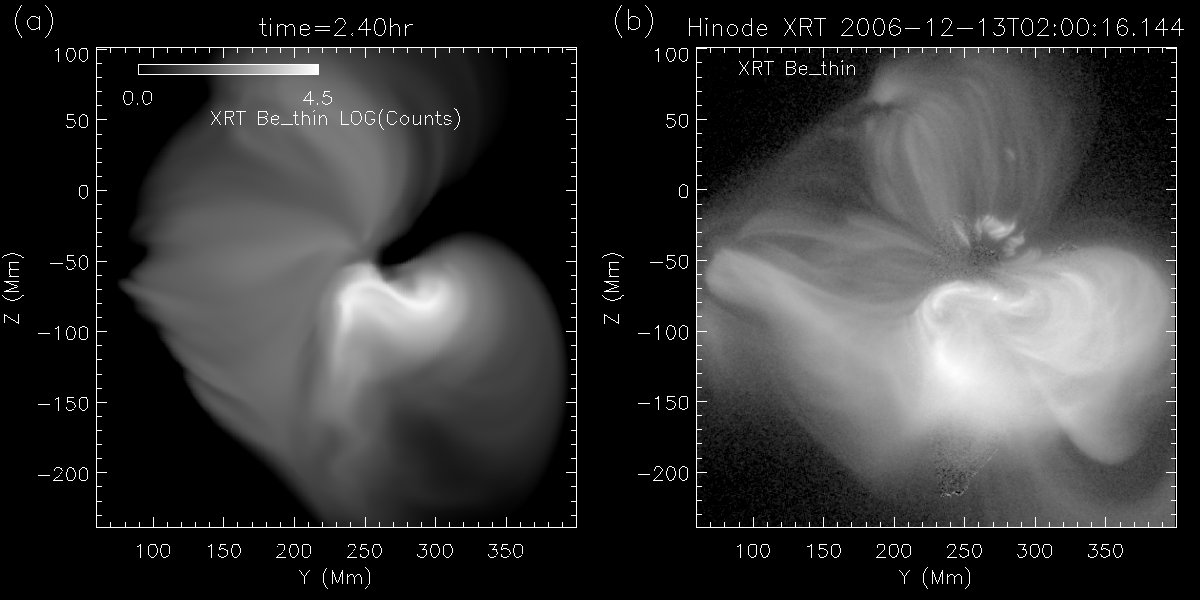}
\caption{(a) Synthetic Hinode/XRT image at the time a few minutes before the onset of the
eruption compared with (b) the observed Hinode/XRT image about 3 minutes before the onset
of the X-class flare.}
\label{fig:compare_xrt}
\end{figure}

\begin{figure}[htb!]
\centering
\includegraphics[width=0.8\textwidth]{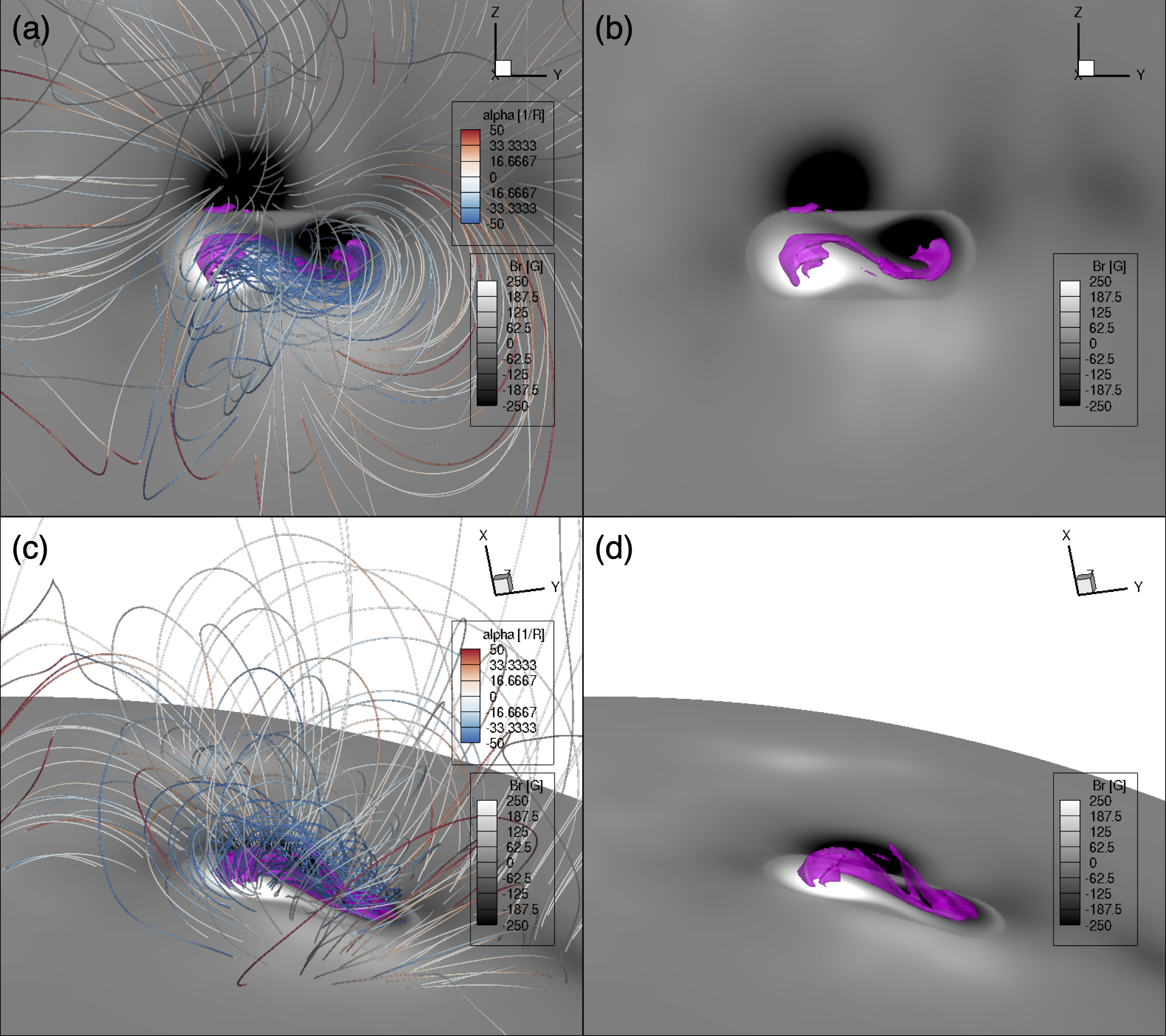}
\caption{Coronal magnetic field lines colored with the twist rate
defined as $\alpha = {\bf J} \cdot {\bf B} / B^2$ and the (purple) iso-surface
of the current density $J = | \nabla \times {\bf B} |$ that outlines the thin current
layer at the time ($t=2.4$ hour) just before the onset of the eruption,
viewed from the earth perspective (upper panels) and a side view (lower panels).}
\label{fig:fdls_jc}
\end{figure}

We find that the sigmoid brightening in the synthetic XRT images sharpens
into a thin sigmoid shape just before (a few minutes before) the onset of the eruption
(Figure \ref{fig:evol_fdl_xrt}(j)). This is consistent with the observed Hinode/XRT image
just before (within a few minutes of) the onset of the X-class flare as shown in a side
by side comparison in Figure \ref{fig:compare_xrt}.
Overall, the synthetic X-ray emission shows qualitatively similar morphology as the
observed Hinode/XRT image for both the ambient coronal loops of the active region
and the central thin sigmoid brightening.
We find that the sharpening of the central sigmoid brightening in the simulation is due
to the resistive heating produced by the formation of a sigmoid shaped strong current
layer in the lower part of the flux rope as shown in Figure \ref{fig:fdls_jc}.
We see that the sigmoid shaped current layer (outlined by the purple iso-surface
of current density $J$) forms in the lower part of the
left-hand twisted (blue) fields of the flux rope, and its central
portion thins into a vertical sheet aligned with the polarity inversion line
(PIL) of the emerging region.
One difference we have noticed between the thin sigmoid in the synthetic XRT
image and that in the observed one is that the modeled sigmoid appears located more
westward by about 10 Mm compared to the observed thin sigmoid. This suggests
that an improvement of the model would be to shift the emerging torus driven at
the lower boundary eastward by about 10 Mm.

\begin{figure}[htb!]
\centering
\includegraphics[width=0.5\textwidth]{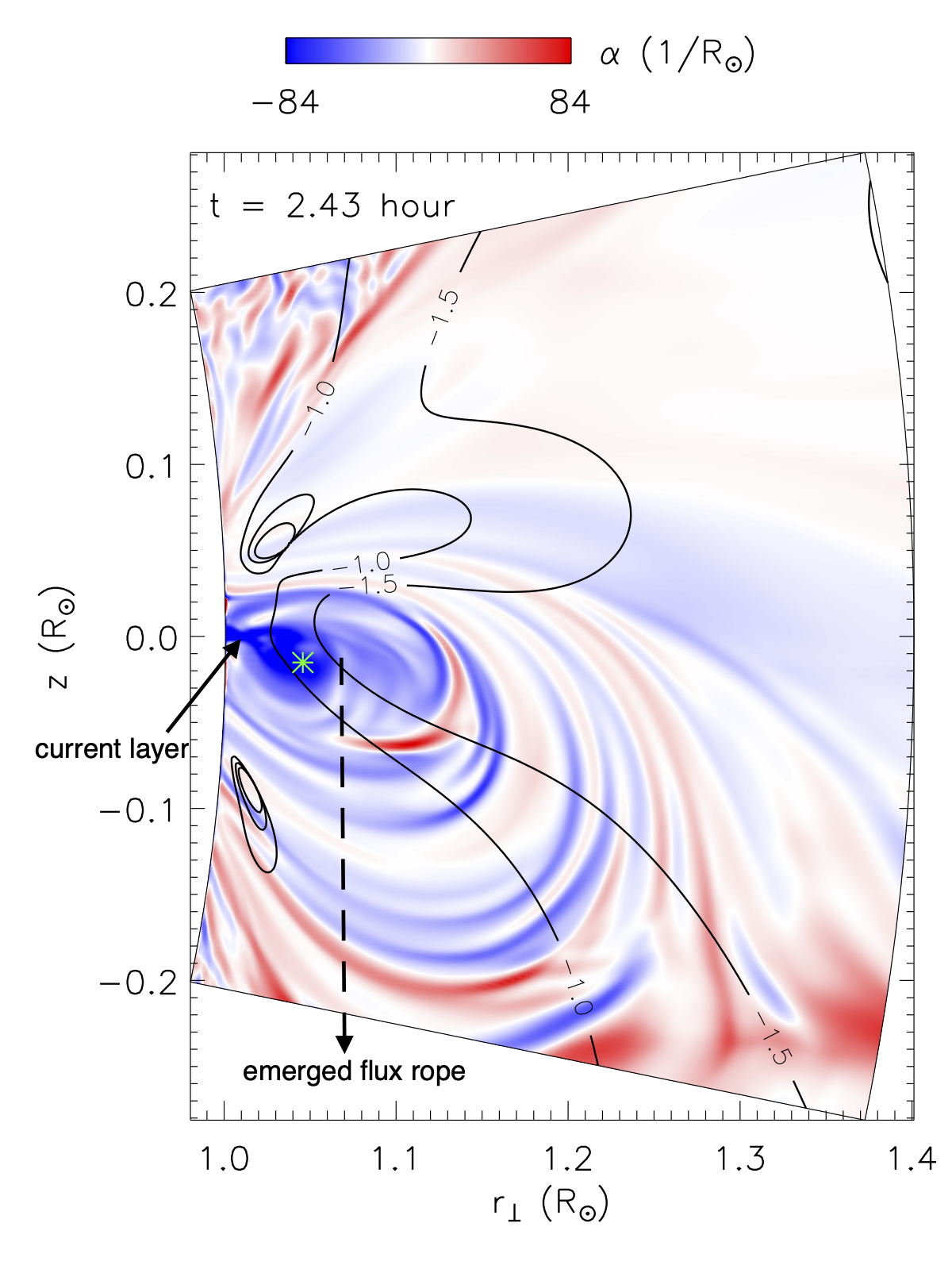}
\caption{Twist rate $\alpha$ in the meridional cross-section across the middle of
the emerged flux rope, over-plotted with the contours of the decay index of the horizontal
component of the corresponding potential field (see text for the decay index definition).
The dashed arrow marks roughly the center of the emerged flux rope cross-section.
The solid arrow points to the thinning current layer near the bottom of the emerged
flux rope. The green star point marks the Lagrangian tracer point,
whose evolution of the radial velocity is shown in Figure \ref{fig:emep_ek_demdt_vr_evol}(e)}
\label{fig:alpha_bpdecay}
\end{figure}

\begin{figure}[htb!]
\centering
\includegraphics[width=0.5\textwidth]{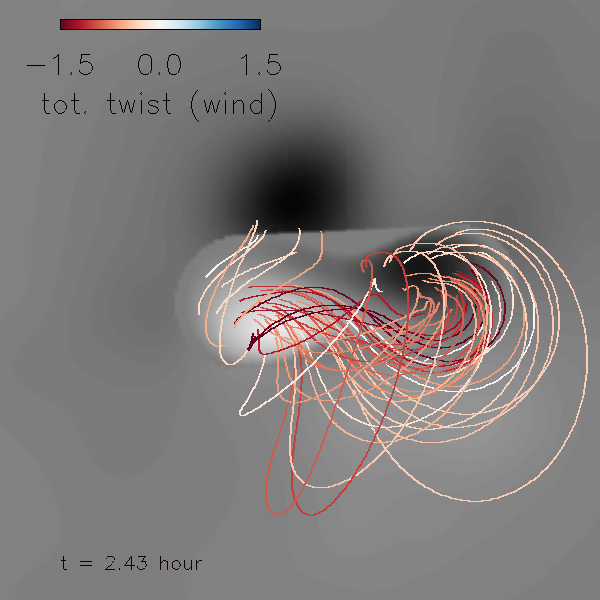}
\caption{The integrated total twist $(1/2 \pi) \int (\alpha/2) ds$ of a set of
field lines in the emerged flux rope (at the same time instance as
Figure \ref{fig:alpha_bpdecay}),
where the integration is along the arc length $s$
of the field line between the two anchored ends at the lower boundary. 
The total twist is shown by the color of the field lines.}
\label{fig:fdls_tottwist}
\end{figure}

Figure \ref{fig:alpha_bpdecay} shows, at the start time ($t=2.43$ hour) of
the exponential acceleration of the flux rope, the twist rate
$\alpha = {\bf J} \cdot {\bf B} / B^2 $ in the meridional
cross-section across the middle of the emerged flux rope, over-plotted with the
contours of the decay index, $d ({\rm ln} P_{\rm h})/d ({\rm ln} h)$,
of the horizontal component of the corresponding potential field $P_{\rm h}$
vs height $h$ above the lower boundary surface.
It can be seen that a significant portion of the emerged flux rope
cross-section (indicated by the dashed arrow) with negative twist rate
has entered above the contour of the critical decay index of -1.5 for the
onset of the torus instability \citep[e.g.][]{Kliem:Toeroek:2006}.
This indicates that the onset of the eruption is consistent with the
onset of the torus instability of the emerged flux rope.
Furthermore, the thinning of the current layer (see the thinning layer of
strong $\alpha$ marked by the solid arrow in Figure \ref{fig:alpha_bpdecay})
suggests the development of a current sheet driven by the unstable rise of
the flux rope, and that the ensuing rapid reconnections enhance the
acceleration of the flux rope.
Figure \ref{fig:fdls_tottwist} shows the integrated total twist
$(1/2 \pi) \int (\alpha/2) ds$ of a set of field lines in the emerged flux
rope, where the integration is along the arc length $s$ of the field line
between the two anchored ends at the lower boundary.
The total twist is shown by the color of these field lines.
We find that for the central core field lines of the emerged flux rope
the total twist has reached about 1.5 winds (or $3 \pi$ radian of rotation)
between the anchored ends,
exceeding the critical total twist of 1.25 winds (or $2.5 \pi$ radian of
rotation) for the onset of the kink instability of a line tied twisted
flux tube \citep{Hood:Priest:1981}.
Thus the kink instability may also have contributed to driving
the unstable rise in the core field region of the flux rope.

\subsection{The erupting magnetic field}
\begin{figure}[htb!]
\centering
\includegraphics[width=1.\textwidth]{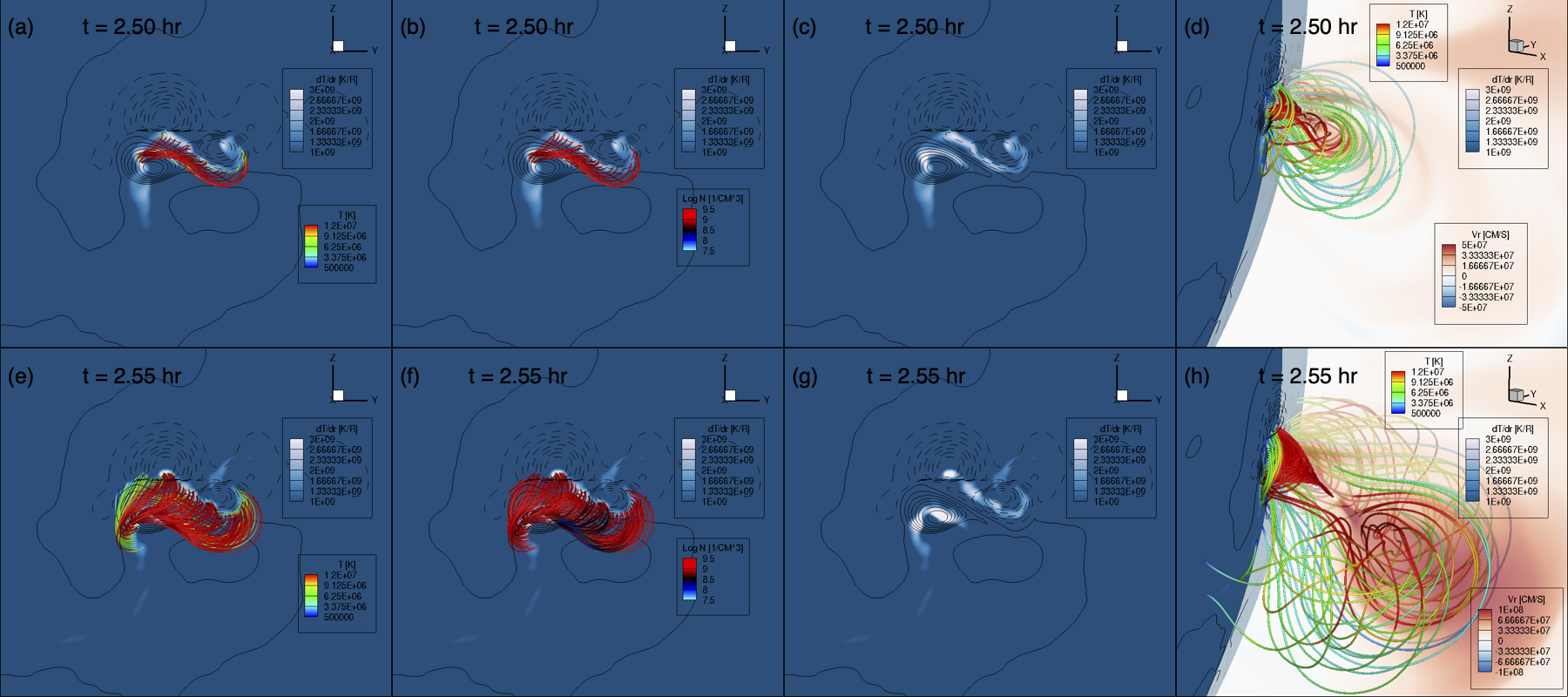}
\caption{(a)(b)(e)(f) show selected field lines of the the post-flare loops colored
with temperature (1st column) and density (2nd column),
plotted against the lower boundary image of the
temperature gradient $dT/dr$ (with high $dT/dr$ identifying the locations of flare
ribbons) and contours of the normal magnetic field $B_r$,
viewed from the earth view point, at two different times (upper row at $t= 2.50$ hour
and lower row at $t=2.55$ hour) during the impulsive phase of the eruption;
(c)(g) are the same as (a)(e) but without the post-flare loop field lines;
(d)(h) are the same as (a)(e) but a different view from the side, and with the addition
of selected field lines of the erupting flux rope and a translucent
plane showing the radial velocity $v_r$ in the central meridional cross-section of
the erupting flux rope. The lower row images for
$t=2.55$ hour correspond to roughly the time of the peak acceleration of
the flux rope and also roughly the time of the peak magnetic
energy dissipation.}
\label{fig:fluxrope_sig_cusp_ribbon}
\end{figure}

During the impulsive phase of the eruption, the central
sharp thin sigmoid in the synthetic X-ray image broadens into
a bright sigmoid shaped band, as can be seen in
Figures \ref{fig:evol_fdl_xrt}(k) and \ref{fig:evol_fdl_xrt}(l) (see
also the associated movie of Figure \ref{fig:evol_fdl_xrt}).
This central brightening corresponds to the highly heated post-flare
(post-reconnection) loops that form due to the rapid reconnection in
the current sheet.
Figure \ref{fig:fluxrope_sig_cusp_ribbon} shows the low-lying field
lines of the post-flare loops as viewed from the earth perspective
(panels (a),(b),(e),(f)) and a side view of the highly twisted field lines
in the erupting flux rope together with the low-lying post-flare
loop field lines (panels (d)(h)), at time $t=2.5$ hour (upper row)
and time $t=2.55$ hour (lower row), which is about the time of the peak
acceleration of the flux rope (Figure \ref{fig:emep_ek_demdt_vr_evol}(e))
and also about the time of the peak dissipation
of the magnetic energy (Figures \ref{fig:emep_ek_demdt_vr_evol}(d)).
In the side view image Figure \ref{fig:fluxrope_sig_cusp_ribbon}(h), one
can see the cusp-shaped tops of the post flare-loops produced by the
reconnection in the current sheet extending above the cusp, and the outflow
$v_r$ from the current sheet as shown in the central cross-section of the
erupting, highly twisted flux rope.
The post-flare loops are heated to high temperature with peak temperature
exceeding $10^7$ K (Figures \ref{fig:fluxrope_sig_cusp_ribbon}(a)(e)) and are
of high density (Figures \ref{fig:fluxrope_sig_cusp_ribbon}(b)(f)). They have
footpoints rooted in the flare ribbons
(Figures \ref{fig:fluxrope_sig_cusp_ribbon}(c)(g)),
corresponding to the region of high temperature gradient $dT/dr$,
with strong downward heat conduction flux.
The evolution of the post-flare loops and the flare ribbons are
consistent with the basic results of the 3D flare reconnection modeled in
many previous MHD simulations of erupting flux ropes
\citep[e.g.][]{Aulanier:etal:2012, Dahlin:etal:2022}. These results
include the formation and separation of two J-shaped flare ribbons from the
PIL (Figures \ref{fig:fluxrope_sig_cusp_ribbon}(c)(g)),
and the post-flare loops that form initially with a stronger shear (the field
component parallel to the PIL) over the
central portion of the PIL (Figures \ref{fig:fluxrope_sig_cusp_ribbon}(a)(b)),
compared to those that form later with a weaker shear
(Figures \ref{fig:fluxrope_sig_cusp_ribbon}(e)(f)).
The formation of sheared post-flare loops and the strong-to-weak shear
transition in time are characteristic features of the 3D flare reconnection 
with a significant guide field resulting from the shear component
of the magnetic field in the erupting flux rope
\citep[e.g.][]{Aulanier:etal:2012, Dahlin:etal:2022}.

\begin{figure}[htb!]
\centering
\includegraphics[width=0.8\textwidth]{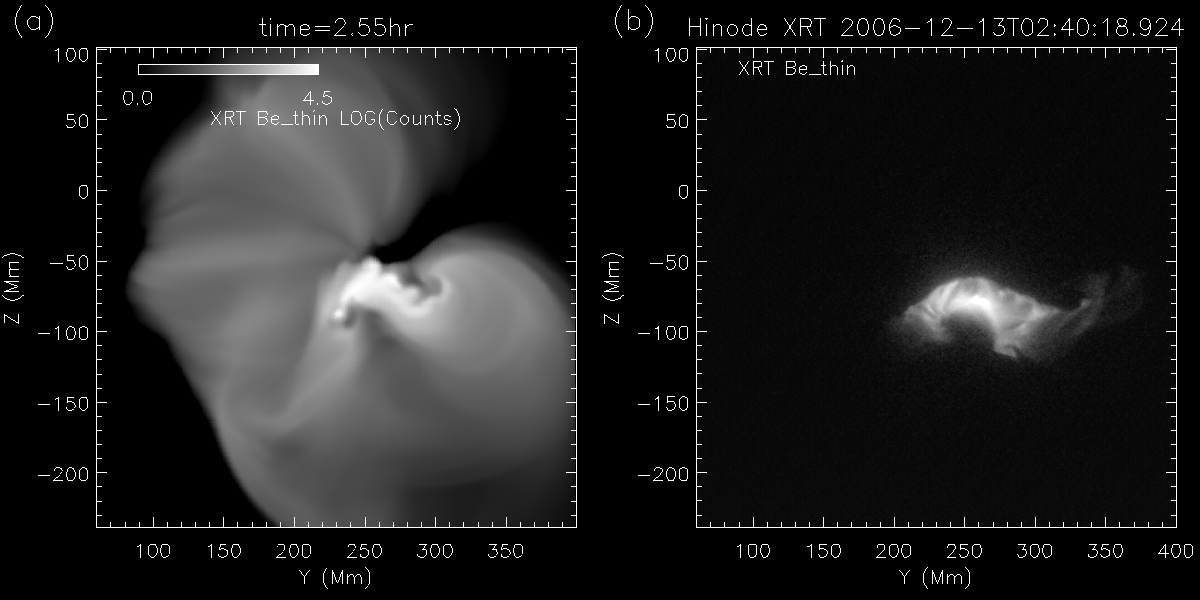}
\caption{(a) Synthetic Hinode/XRT image at the time of approximately the peak of
the magnetic energy dissipation compared to (b) the observed Hinode/XRT image
at the time of peaking soft X-ray emission \citep{Schrijver:etal:2008}.}
\label{fig:compare_xrt_impulse}
\end{figure}

The high-temperature, high-density post-flare loops
(e.g. Figures \ref{fig:fluxrope_sig_cusp_ribbon}(e)(f))
produce the central enhanced emission of a sigmoid
shaped band seen in the synthetic X-ray image (e.g.
Figure \ref{fig:compare_xrt_impulse}(a)).
In comparison, the observed XRT image (Figure \ref{fig:compare_xrt_impulse}(b))
at the impulsive phase also shows a qualitatively similar sigmoid shaped band
of post-flare loops, although the band appears more extended towards both
the east and west ends.
Note that in the observed XRT image during the impulsive phase, only the
central post-flare loops are visible.  This is because during the flare phase
the soft X-ray brightening produced by the flare loops is so much higher than
that from the surrounding non-flare loops that a much shorter exposure time
was used to avoid saturation such that the ambient loops become invisible.
On the other hand, the modeled synthetic X-ray images show much less
enhancement in the emission from the post-reconnection loops relative to
the ambient coronal loops, and thus both remain visible in the dynamic
range of the brightness displayed.  Here the comparison between the modeled
images and the observed ones are merely regarding the qualitative
morphological features and is not quantitative.
The comparison (see Figure \ref{fig:compare_xrt_impulse})
shows that the observed sigmoid shaped band of post-flare loops
appears more extended in the east-west
direction and shows a less curved hook at its west end.
In our simulation, we imposed the emergence of an idealized magnetic
torus which only roughly mimics the observed flux emergence pattern
(see Figures \ref{fig:init}(b) and \ref{fig:init}(d)).
The resulting emerging bipolar region at the time of the onset
of the eruption does not have as large a polarity separation as the observed
case, and for the leading (negative) polarity at the west side, the observed
flux distribution is more fragmented and extended. These detailed
differences can all be contributing to differences in the extension and
shape of the current sheet that forms and hence the differences in
the extension and shape of the band of the post flare loops that develop.

During the impulsive phase of the eruption, we also see in the movie
of the synthetic XRT emission assocated with Figure \ref{fig:evol_fdl_xrt},
an outgoing band of diffuse brightening that emanates from the sigmoid and
moves southward. This outgoing diffuse brightening corresponds to the hot
core of the erupting flux rope, whose temperature reaches as high as about
$10^7$ K, as can be seen in Figures \ref{fig:fluxrope_sig_cusp_ribbon}(d)(h)
which show the erupting flux rope field lines colored with temperature.
Such an outgoing brightening feature in soft X-ray produced by an
erupting flux rope has been discussed in \citet{McKenzie:Canfield:2008}.
In the Hinode/XRT observation of the present event, the brightening from
the post-flare loops dominates and saturates the images
(Figure \ref{fig:compare_xrt_impulse}(b)) and any
disturbances outside of the flaring region are not seen.

\begin{figure}[htb!]
\centering
\includegraphics[width=0.8\textwidth]{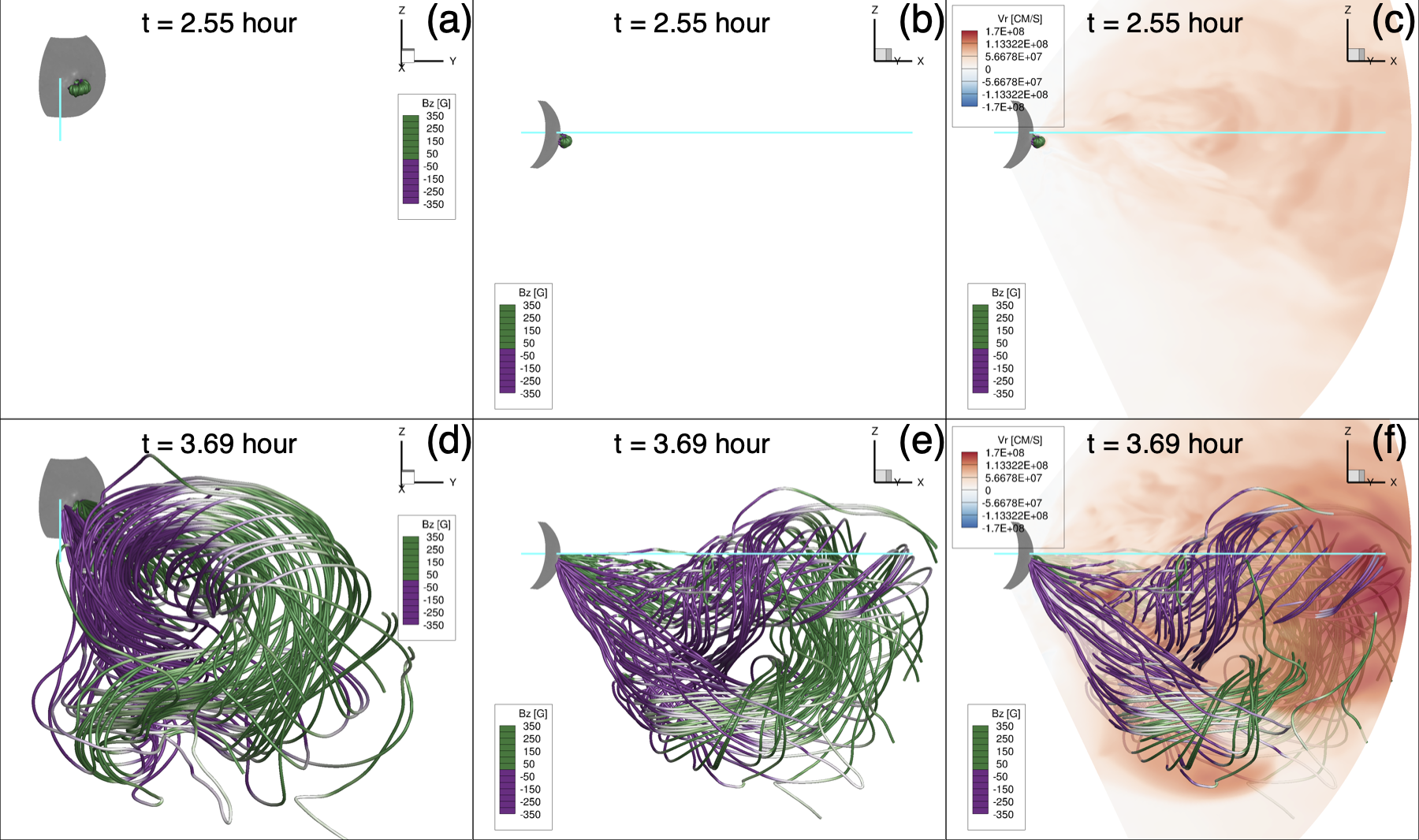}
\caption{3D views of selected field lines of the erupting flux rope at
$t=2.55$ hour (top row) and at $t=3.69$ hour (bottom row), and viewed from
an on-disk view point (left column) and from a east-limb view point
(middle column). The right column is the same as the middle column but with
a middle meridional cross-section across the flux rope added showing the
radial velocity. The field lines are colored with the sign of $B_z$.
The cyan line denotes the sun-earth line. An animated version of the images
in the left and middle columns are available, which shows the 3D dynamic
evolution of the erupting flux rope from $t=2.4$ hour (just before the onset
of the eruption) to $t=3.69$ hour (when the flux rope begins to exit the domain).}
\label{fig:eruptrope}
\end{figure}

Figure \ref{fig:eruptrope} shows selected field lines of the erupting flux rope
at two time instances during the eruption (top row at $t=2.55$ hour at about the peak
acceleration of the flux rope, and bottom row at $t=3.69$ hour, when the flux rope
begins to exit the domain) and an associated animation which shows the 3D dynamic
evolution of the erupting flux rope.
The selected field lines are traced from a set of
Lagrangian tracers originally placed in the emerged flux rope at the
pre-eruption phase (at $t=2.4$ hour), and tracked in the subsequent velocity
field.
It can be seen that the initially nearly east-west oriented flux rope has
rotated by about $90^{\circ}$ counter-clockwise to the nearly south-north
oriented final erupting flux rope as shown in
Figures \ref{fig:eruptrope}(d) and \ref{fig:eruptrope}(e).
The initial west (east) leg of the flux rope rotates to become the
north (south) leg of the final flux rope exiting the domain.
The flank of the north leg of the erupting flux rope with south-ward
directed $B_z$ (purple field lines) is adjacent to the sun-earth line,
although not yet intersecting the sun-earth line
(see Figures \ref{fig:eruptrope}(d) and \ref{fig:eruptrope}(e)).
The ambient solar wind condition strongly affects the kinematics
of the erupting flux rope.
Because of the fast wind outflow north of
the emerging flux rope as can be seen in Figure \ref{fig:eruptrope}(c),
the northern portion of the erupting flux rope erupts into the fast wind
stream, reaching a higher outward speed approaching $\sim 1700$ km/s
(Figure \ref{fig:eruptrope}(f)) and extends out further in the front
(Figures \ref{fig:eruptrope}(d) and \ref{fig:eruptrope}(e)).
The rotation of the erupting flux rope found here (about $90^{\circ}$)
is significantly different from that found in \citet{Fan:2016}, where a
much greater rotation of nearly $180^{\circ}$ was found. 
This is caused by both the differences in the imposed lower boundary
flux emergence as well as the presence of an ambient solar wind with a
partially open magnetic field in this simulation instead of a static,
potential field background condition assumed in \citet{Fan:2016}.

\section{Summary and discussion}
\label{sec:summary}
In this work, we have carried out a new simulation of the
13-December-2016 CME event from AR 10930, improving upon the previous
simulation by \citet{Fan:2016}.  The improvements include the following.
Instead of using a potential magnetic field with a static polytropic
atmosphere as the initial state, we have initialized a partially open
coronal magnetic field with a background solar
wind as the initial state as described in section \ref{sec:model}.
For the normal magnetic flux distribution at the
lower boundary (the base of the corona) of this initial
state, we use the normal magnetic field in the horizontal plane at 14 Mm height
of the potential field extrapolated from the observed photosphere
magnetogram, instead of heavily smoothing the photosphere magnetogram
as was done in \citet{Fan:2016}.
The resulting lower boundary normal flux distribution
contains a significantly stronger peak field strength ($\sim 500$ G)
compared to that in \citet{Fan:2016}.
Furthermore, the current simulation uses a more explicit treatment of the
background coronal heating. We impose at the lower boundary a random electric
field representing the effect of turbulent convection that drives field-line
braiding, and the consequent (numerical) resistive and viscous heating
provides the background coronal heating and accelerates the solar wind.
Together with this random electric field, we also impose the emergence
of a twisted magnetic torus in the way similar to \citet{Fan:2016}, but is
more compact and of a stronger field strength to build up the non-potential,
twisted pre-eruption coronal magnetic field of the active region.

With the inclusion of the random electric field which drives a
coronal heating that varies spatially and temporally with the formation
of the strong current layers in the given 3D coronal magnetic field,
we are able to model the synthetic soft X-ray images that
would be observed by the Hinode/XRT.
It is found that the simulated pre-eruption active
region magnetic field, where a twisted magnetic flux rope is built up
by the imposed flux emergence, produces synthetic soft X-ray emission
that shows qualitatively similar morphology as that observed by the
Hinode/XRT, for both the ambient coronal loops and the central
sigmoid that sharpens just before the onset of the eruption.
The sharpening of the central sigmoid is due to the formation of a
thin sigmoid shaped current layer in the lower part of the flux rope
along the central PIL.
It is found that at the onset of the exponential acceleration of
the flux rope, a significant portion of the emerged flux rope cross-section
has entered above the contour of the critical decay index of -1.5 for the
onset of the torus instability \citep[e.g.][]{Kliem:Toeroek:2006}. Thus
the onset of the eruption is consistent with the onset of the torus
instability. At the same time we also find that for the central
core field lines of the emerged flux rope, the total twist has reached about
1.5 winds, exceeding the critical total twist of 1.25 winds for the onset
of the kink instability \citep[e.g.][]{Hood:Priest:1981}. This suggests
that both the torus instability and the kink instability contribute to
driving an unstable rise of the flux rope, which in turn drive the
thinning of the central current layer, and the resulting rapid
reconnection further leads to the impulsive acceleration of the flux
rope.

We find that during the impulsive phase of the eruption, the central sharp
thin sigmoid in the synthetic X-ray image broadens to become a bright
sigmoid-shaped band, which corresponds to the row of low-lying
high-temperature, high-density post-flare loops that form due to the rapid
reconnection in the current sheet.
The Hinode/XRT observation shows a similar evolution of transitioning from
the thin sigmoid to a broadened sigmoid-shaped band of post-flare loops.
The agreement is only qualitative. The thin sigmoid at the onset of the
eruption appears more westward (by about 10 Mm) in the synthetic image
compared to the observed one. In the impulsive phase, the sigmoid-shaped
band of post-flare loop brightening in the synthetic XRT image appears less
extended towards both the east and west ends,
and shows a more curved hook at the west end.
Our simulation imposed the emergence of an idealized magnetic torus
which only roughly mimics the observed flux emergence pattern.
The above discrepancies in the morphology of the soft X-ray emission
suggest that a better observationally constrained lower
boundary driving of flux emergence is needed to better reproduce the flux
rope and the current sheet that form in the realistic coronal magnetic field
of the event.

We find that the erupting flux rope accelerates to a peak speed of about
$1700$ km/s. The erupting flux rope rotates by about $90^{\circ}$
counter-clock wise, from an initially nearly east-west oriented emerged
flux rope to a final nearly north-south oriented flux rope exiting the
domain. The west leg of the initial flux rope rotates to become the
north leg of the erupting flux rope, and the flank of the north leg of the
flux rope with south-ward directed Bz is adjacent to the sun-earth line,
although not yet intersecting the sun-earth line, when the flux rope is
exiting the domain. The rotation of the erupting flux rope found here
is significantly smaller than that found in \citet{Fan:2016}, where
a rotation of nearly $180^{\circ}$ was obtained. This is caused by
both the difference in the lower boundary flux emergence as well as
the difference in the ambient magnetic field and solar wind conditions.
We find that the ambient solar wind condition significantly affects
the kinematics of the erupting flux rope. In this case the north
portion of the erupting flux rope erupts into the adjacent fast
solar wind stream and hence attains a higher erupting speed and
extends out further in the front.
Thus a better observationally constrained lower boundary
driving condition for the emerging active region and also correctly
determining the conditions of the ambient coronal magnetic field and
the solar wind are all needed to determine the realistic kinematics and
the magnetic structure of the outgoing CME.

\acknowledgments
The author thanks the anonymous referee for helpful comments.
This material is based upon work supported by the National Center for Atmospheric
Research, which is a major facility sponsored by the National Science Foundation
under Cooperative Agreement No. 1852977. 
This work is also supported by the NASA LWS grant 80NSSC19K0070 to NCAR.
The author would like to acknowledge high-performance computing support from Cheyenne
(doi:10.5065/D6RX99HX) provided by NCAR's Computational and Information Systems
Laboratory, sponsored by the National Science Foundation.
Resources supporting this work were also provided by the NASA High-End Computing (HEC)
Program through the NASA Advanced Supercomputing (NAS) Division at Ames Research Center.

\end{document}